\documentclass[12pt]{iopart}

\newcommand\ee{\end{equation}}
\newcommand\be{\begin{equation}}
\newcommand\eea{\end{eqnarray}}
\newcommand\bea{\begin{eqnarray}}

\newcommand{\F}{\textrm{F}}

\newcommand{\bn}{\hat{\mathbf{n}}}
\newcommand{\bN}{\hat{\mathbf{N}}}
\newcommand{\bV}{\mathbf{V}} 
\newcommand{\bv}{\mathbf{v}}
\newcommand{\bx}{\mathbf{x}}

\newcommand{\bk}{\mathbf{k}}

\newcommand{\HH}{{\cal H}}

\newcommand{\de}{\delta}
\newcommand{\dd}{\partial}

\usepackage{iopams}  
\usepackage{epsfig}
\usepackage{graphicx}

\begin{document}

\title[]{Isolating relativistic effects in large-scale structure}

\author{Camille Bonvin}

\address{Kavli Institute for Cosmology Cambridge and Institute of Astronomy, Madingley Road, Cambridge CB3 OHA, U.K.
\&
DAMTP, Centre
for Mathematical Sciences, Wilberforce Road, Cambridge CB3 0WA, U.K.}
\ead{cbonvin@ast.cam.ac.uk}
\begin{abstract}
We present a fully relativistic calculation of the observed galaxy number counts in the linear regime. We show that besides the density fluctuations and redshift-space distortions, various relativistic effects contribute to observations at large scales. These effects all have the same physical origin: they result from the fact that our coordinate system, namely the galaxy redshift and the incoming photons' direction, is distorted by inhomogeneities in our universe. We then discuss the impact of the relativistic effects on the angular power spectrum and on the two-point correlation function in configuration space. We show that the latter is very well adapted to isolate the relativistic effects since it naturally makes use of the symmetries of the different contributions. In particular, we discuss how the Doppler effect and the gravitational redshift distortions can be isolated by looking for a dipole in the cross-correlation function between a bright and a faint population of galaxies.    
\end{abstract}


\section{Introduction}

Galaxy surveys are extremely useful to probe fundamental properties of our universe. Inhomogeneities in the distribution of galaxies can indeed be related to the underlying inhomogeneous distribution of dark matter. By observing fluctuations in the galaxy distribution at different redshifts, one can therefore study the growth of dark matter perturbations and test the theory of gravitation. 

In 1987, Kaiser extended the interest and utility of galaxy surveys, by showing that the observed distribution of galaxies is not only sensitive to the dark matter distribution, but also to the peculiar velocity of galaxies~\cite{Kaiser_rsd}. His argument was based on the fact that observations are made in redshift-space, and that peculiar velocities affect the redshift via the Doppler effect. This so-called {\it redshift-space distortion} has proven to be rich in information, allowing at the same time to measure the large-scale velocity field and to alleviate the problem of the bias.

In addition, Broadhurst et al.~\cite{broadhurst} and Moessner et al.~\cite{moessner} showed that gravitational lensing also affects the observed distribution of galaxies, via the {\it magnification bias}. This effect relies on the fact that surveys are limited in magnitude and that gravitational lensing changes the observed luminosity of galaxies, allowing galaxies that would be too faint to be detected to be magnified and consequently to pass the threshold of observation. Through the magnification bias, the distribution of galaxies becomes sensitive to the sum of the two metric potentials, integrated along the photons' trajectory. 

These three contributions to the observed fluctuations of galaxies, namely the dark matter density fluctuations, the redshift-space distortions and the magnification bias, show the potential of galaxy surveys to probe the content and the geometry of our universe. During the last years, it has however been shown that these three contributions represent only some of the many physical distortions that affect the observed distribution of galaxies~\cite{Yoo, pap_gal, challinor}. A systematic derivation of the galaxy number counts shows how perturbations in the density, the velocity and the metric distort our observed coordinate system (i.e. the galaxy redshift $z$ and the direction of incoming photons $\bn$) and subsequently affect our observables. 

These novel distortions, which reflect the effect of general relativity onto the photons' propagation,  are suppressed at subhorizon scales with respect to the standard redshift-space contributions. Their impact on the observed galaxy number counts is therefore more subtle and difficult to detect. However, the potential of these {\it relativistic effects} for testing the consistency of general relativity is important~\cite{lombriser}, and justifies the search for new observables, tailored to detect and to isolate them.

In this paper, we review the relativistic derivation of the observed galaxy number counts at linear order, presented in~\cite{pap_gal} and~\cite{pap_conv}. We discuss the physical meaning of the new distortions and compare them with the standard expression derived by Kaiser~\cite{Kaiser_rsd}. We then link this calculation with observable quantities by showing the impact of the relativistic effects on the angular power spectrum~\cite{pap_gal} and on the two-point correlation function in configuration space~\cite{pap_asymm}. We discuss and compare the relative advantages and drawbacks of these two statistical measures and we show how the relativistic effects can be isolated by looking for a dipole in the correlation function between two populations of galaxies.      

\section{The relativistic number density of galaxies}

\subsection{General expression}

In a galaxy survey, one can measure the fluctuations in the number of galaxies across the sky. Galaxy maps can be pixelised in bins of redshift and solid angle. Let us call $N(z,\bn)$ the number of galaxies detected in a pixel centred at $(z,\bn)$ (here $\bn$ denotes the direction of observation) of size $dz$ and solid angle $d\Omega$. At each redshift one can average the number of galaxies over all directions $\bn$ to find the mean number $\bar N(z)$. The galaxy fractional number overdensity is then simply defined as 
\be
\label{Delta}
\Delta(z,\bn)=\frac{N(z,\bn)-\bar{N}(z)}{\bar{N}(z)}\, .
\ee
We want to relate $\Delta$ to the dark matter density fluctuations. 
We express the number of galaxies as $N(z,\bn)=\rho(z,\bn)\cdot V(z,\bn)$, where $\rho$ is the galaxy number density and $V$ denotes the volume of the pixel. Expanding both $\rho$ and $V$ around their mean value and keeping only linear terms, equation~(\ref{Delta}) becomes
\be
\Delta(z,\bn)=\frac{\rho(z,\bn)-\bar\rho(z)}{\bar\rho(z)}+\frac{\delta V(z,\bn)}{\bar V(z)}\, . \label{deltaz}
\ee
The first term in equation~(\ref{deltaz}) can be related to the dark matter density contrast
\be
\delta(z,\bn)\equiv\frac{\rho(z,\bn)-\bar\rho(\bar z)}{\bar\rho(\bar z)}\, ,
\ee
by Taylor-expanding $\bar\rho$ around the background redshift $\bar z$
\be
\bar\rho(z)=\bar{\rho}(\bar z+\delta z)\simeq\bar\rho(\bar z)+\partial_z\bar\rho\cdot \delta z\, .
\ee
With this we obtain
\be
\label{Delta_gen}
\Delta(z,\bn)= \delta(z,\bn)-3\frac{\delta z}{1+\bar z}+\frac{\delta V(z,\bn)}{\bar V}\, .
\ee
Equation~(\ref{Delta_gen}) shows that the observed fluctuations in the number density of galaxies is not only affected by the underlying dark matter fluctuations, but also by fluctuations in the observed redshift and in the observed volume of each pixel. 

Well inside the horizon, equation~(\ref{Delta_gen}) is dominated by two terms: the density fluctuations, $\delta$, and the distortions induced by galaxy peculiar velocities on the volume of the pixel, $\delta V/\bar V$. 
This combination of terms gives rise to the well-known redshift-space distortions expression~\cite{Kaiser_rsd}
\be
\Delta(z,\bn)=\delta(z,\bn)-\frac{1}{\mathcal{H}}\partial_r(\mathbf{v}\cdot\bn)\, , \label{Delta_Newt}
\ee
where $r$ denotes the radial comoving distance, $\HH$ is the comoving Hubble parameter, and $\bv$ is the galaxy peculiar velocity. In his original paper~\cite{Kaiser_rsd}, Kaiser showed that in addition to these dominant terms, the mapping from real to redshift space generates subdominant contributions, proportional to $\bv\cdot \bn$
\be
\Delta(z,\bn)=\delta(z,\bn)-\frac{1}{\mathcal{H}}\partial_r(\mathbf{v}\cdot\bn) -\frac{1}{\HH r}\left( 2+\frac{d\ln \phi}{d\ln r}\right)\,\mathbf{v}\cdot \bn\, , \label{Delta_Kaiser}
\ee
where $\phi$ is the selection function. Using Einstein's equations, we can compare the relative importance of the three contributions. Poisson equation tells us that $\delta\propto (k/\HH)^2\Phi$, and the (0i)-Einstein's equation implies that $v\propto (k/\HH)\Psi$. The first two terms are therefore enhanced by one factor $k/\HH$ with respect to the last term (note that in the last term $r\HH\sim 1$). For modes well inside the horizon, $k\gg \HH$, the last term is negligible and we recover expression~(\ref{Delta_Newt}). At large scales however, near the horizon, the last term cannot be neglected.

Expression~(\ref{Delta_Kaiser}) has been derived in a Newtonian way and it contains therefore only some of the redshift and volume corrections present in (\ref{Delta_gen}). This expression also has the problem of being gauge-dependent (which should not be the case for an observable quantity) since both the density $\delta$ and the velocity $\bv$ are gauge-dependent. To consistently include all the physical distortions contributing to $\Delta$ and to solve the problem of gauge-dependence,  it is necessary to perform a relativistic calculation of the volume fluctuations and of the redshift fluctuations. As we will see, the fully relativistic expression contains a number of contributions that are proportional to $(k/\HH)\Psi$, i.e. of the same order as the last term in~(\ref{Delta_Kaiser}), as well as contributions that are directly proportional to $\Psi$, i.e. suppressed by one additional factor of $\HH/k$. 

\subsection{Derivation}

We summarise here the relativistic derivation of $\Delta$ in a perturbed Friedmann universe. As explicitly shown in~\cite{pap_gal}, the result is gauge-independent. We can therefore express $\Delta$ in terms of gauge invariant variables only, namely the two Bardeen potentials $\Phi$ and $\Psi$~\footnote{Our convention for the Bardeen potentials is such that in the Newtonian gauge the metric reads $ds^2=-a^2\big(1+2\Psi\big)d\eta^2+a^2\big(1-2\Phi\big)\delta_{ij}dx^idx^j$.}, and the gauge invariant density and velocity
variables. 

The redshift fluctuations can be calculated by solving the null geodesic equation and calculating the change in energy between emission and observation. The redshift perturbation itself, $\delta z$, is not gauge-independent since the splitting between the background and the perturbation depends on the gauge choice. The combination of the first two terms in equation~(\ref{Delta_gen}) is however gauge-independent. It reads
\be
\de_z(z, \bn)\equiv\delta-3\frac{\delta z}{1+z}= D^{({\rm g})}_s -3\bV\cdot\bn +3\Psi+3\int_{0}^{r} dr' (\dot{\Phi}+\dot{\Psi})\, ,\label{dez}
\ee
where $D^{({\rm g})}_s$ and $V$ are the gauge-invariant variables describing respectively the galaxy density and the peculiar velocity in the longitudinal gauge. The relation between the galaxy number density and the dark matter density contrast is well defined in the comoving gauge~\cite{bruni, biasbaldauf, biasjeong}. In the simplest case, one assumes a linear bias~\cite{bias}, $D^{({\rm g})}=b\cdot D$, where $D$ denotes the gauge-invariant density in the comoving gauge, so that in Fourier space
\be
D_s^{({\rm g})}=b\cdot D-3\frac{\HH}{k}V\, .
\ee

To calculate the volume fluctuations, $\delta V/\bar V$, one needs to map the physical volume of the pixel 
\be
dV=\sqrt{-g}\;\epsilon_{\mu\nu\alpha\beta}\;u^\mu {\rm d}x^\nu {\rm d}x^\alpha {\rm d}x^\beta \, ,
\ee
with the observed volume $dz\cdot d\Omega_O$. Here $u^\mu$ denotes the 4-velocity of the galaxies inside the infinitesimal volume $dV$. The coordinates at the source position ${\rm d}x^\alpha$ can be related to the observed coordinates $z$, $\theta_O$ and $\varphi_O$ through
\be
dV =\sqrt{-g}\;\epsilon_{\mu\nu\alpha\beta}u^\mu\! \frac{\dd x^\nu}{\dd z} \! 
\frac{\dd x^\alpha}{\dd \theta_S}\! \frac{\dd x^\beta}{\dd \varphi_S}\! 
\left| \frac{\dd (\theta_S,\varphi_S)}{\dd (\theta_O,\varphi_O)} 
\right|\! {\rm d}z {\rm d}\theta_O{\rm d}\varphi_O \, .  \label{vol}
\ee
Here the angles $\theta_S$ and $\varphi_S$ represent the photon's direction at emission. 
In a homogeneous and isotropic Friedmann universe, photons propagate on straight lines, $\theta_S=\theta_O$ and $\varphi_S=\varphi_O$, and the observed volume of the pixel is independent of the direction of observation $\bn$
\be
d\bar V(\bar z)=\frac{\bar r^2}{(1+\bar z)^4\HH}\,d\bar z \sin\theta_Od\theta_Od \varphi_O\, .
\ee
In a perturbed universe however, both the photon's direction and the length of the geodesic are perturbed
\be
\theta_S=\theta_O+\delta\theta, \quad \varphi_S=\varphi_O+\delta\varphi \quad {\rm and} \quad r=\bar r+\delta r\, .
\ee 
These fluctuations translate directly into fluctuations in the volume
\bea
\fl \frac{\de V}{\bar V}(z, \bn) =    
 -3\Phi+\left( \cot\theta_O+\frac{\dd}{\dd \theta}\right)\de \theta +
\frac{\dd \de \varphi}{\dd \varphi} + \bV\cdot \bn+\frac{2\de r}{r}
-\frac{d\de r}{d\lambda}\nonumber\\
+\frac{1}{\HH(1+ z)}\frac{d \de z}{d\lambda}
-\left(\frac{2}{ r\HH}+\frac{\dot{\HH}}{\HH^2}-4 \right)\frac{\de z}{1+ z}\, ,
\eea
where $\lambda$ is the affine parameter of the photon geodesic. The fluctuations $\delta r, \delta \theta$ and $\delta\varphi$ can be explicitely calculated by solving the null geodesic equation in a perturbed Friedmann universe. The volume perturbations become
\bea
\label{vol_expl}
\fl\frac{\de V}{\bar V}=-2(\Psi+\Phi) + 4\bV\cdot\bn +\frac{1}{\HH}
\left[\dot\Phi+\dd_r\Psi-\frac{d(\bV\cdot\bn)}{dr}\right] \\
+\left(\frac{\dot{\HH}}{\HH^2}+\frac{2}{r\HH}\right)
\left[\Psi-\bV\cdot\bn+ \int_{0}^{r} dr'(\dot{\Phi}+\dot{\Psi})\right]
-3\int_{0}^{r} dr'(\dot{\Phi}+\dot{\Psi})\nonumber\\
+ \frac{2}{r}\int_{0}^{r} dr' (\Phi+\Psi)
- \frac{1}{r}\int_{0}^{r} dr'\frac{r-r'}{r'}
\Delta_\Omega(\Phi+\Psi)~, \nonumber 
\eea
where $\Delta_\Omega$ denotes the Laplacian transverse to the line-of-sight.

Combining equation~(\ref{dez}) with equation~(\ref{vol_expl}), the observed overdensity of galaxies becomes
\bea
\fl \Delta(z,\bn) = D^{({\rm g})}_s -\frac{1}{\HH} \dd_r(\bV\cdot\bn)+\frac{1}{\HH}\dot{\bV}\cdot\bn
+\left(1-\frac{\dot{\HH}}{\HH^2}-\frac{2}{r\HH} \right)\bV\cdot\bn +\frac{1}{\HH}\dd_r\Psi\nonumber\\
 - \frac{1}{r}\int_{0}^{r} dr'\,\frac{r-r'}{r'} \Delta_\Omega(\Phi+\Psi)+\left(\frac{\dot{\HH}}{\HH^2}+\frac{2}{r\HH} \right)\left[\Psi+\int_0^{r}dr'(\dot{\Phi}+\dot{\Psi})\right]\nonumber\\
+\frac{2}{r}\int_{0}^{r} dr' (\Phi+\Psi) +\Psi-2\Phi+\frac{1}{\HH}\dot{\Phi}\, . \label{Delta_pap}
\eea
The first term in equation~(\ref{Delta_pap}) describes the true fluctuations in the distribution of galaxies, related to the fact that the dark matter is inhomogeneously distributed. All the other terms represent distortions in the coordinate system in which we are making our observations, namely the galaxy redshift and the incoming direction of the photons.  

Comparing expression~(\ref{Delta_pap}) with its Newtonian counter-part~(\ref{Delta_Kaiser}), we see first that the velocity terms are different. The third, fourth and fifth terms in the first line of~(\ref{Delta_pap}) are not present in~(\ref{Delta_Kaiser}). They describe the fact that the mapping between real and redshift-space does not only affect the comoving distance of galaxies, $r$, but also the comoving time at which the observed photons are emitted, since we observe on our past light-cone, $\eta=\eta_0-r$. The third term accounts for the fact that a wrong estimation of the photons emission time translates into a wrong estimation of the galaxy velocity at the time of emission, since velocities evolve with time. The fifth term accounts for the fact that a wrong estimation of the photons emission time translates into a wrong estimation of the background expansion rate of the universe at the time of emission, since the Hubble parameter $\HH$ evolves with time. And the fourth term, which has been called the {\it light-cone effect} in~\cite{Kaiser_redshift, pap_asymm}, describes the fact that by changing the size of the redshift bin, the peculiar velocity of galaxies also changes the size of the time interval $d\eta$. 

In addition to the different velocity terms, the first line in~(\ref{Delta_pap}) contains also a term proportional to the gradient of the gravitational potential $\dd_r\Psi/\HH$. This contribution describes the change in the comoving size of the redshift bin due to gravitational redshift. It is therefore the direct equivalent of the standard redshift distortions term, $\dd_r(\bV\cdot\bn)/\HH$, and we call it {\it gravitational redshift distortions}. Due to the gradient of $\Psi$, this term is effectively of the same order of magnitude as the velocity terms.  In theories of gravity that obey Euler equation
\be
\dot{\bV}\cdot\bn+\HH \bV\cdot\bn+\dd_r\Psi=0\, , \label{Euler}
\ee
the corrections in the first line of~(\ref{Delta_pap}) combine to
\be
-\left(\frac{\dot{\HH}}{\HH^2}+\frac{2}{r\HH} \right)\bV\cdot\bn\, ,
\ee
which differs from expression~(\ref{Delta_Kaiser}) by the term $\dot \HH/\HH^2$ only~\footnote{We will discuss below the contribution from the selection function $\phi$.}.

The second and third lines of~(\ref{Delta_pap}) have no Newtonian counter-parts in~(\ref{Delta_Kaiser}). The first term in the second line of~(\ref{Delta_pap}) describes the distortions generated by gravitational lensing on the solid angle of observation $d\Omega$. As we will see below, this term combines with magnification bias to give rise to the standard lensing expression derived in~\cite{broadhurst, moessner}. This contribution is sensitive to the Laplacian of the gravitational potentials along the line-of-sight. At first sight it seems to be of the same order of magnitude as the density term (Poisson equation tells us indeed that $(k/\HH)^2 \Phi\sim D$). However, the Laplacian $\Delta_\Omega$ is a transverse Laplacian, which is effectively two-dimensional and therefore brings in only one power of $k$, namely $k_\perp/\HH\,(\Phi+\Psi)$, where $\bk_\perp$ is orthogonal to $\bn$. The gravitational lensing term is therefore suppressed by one power of $\HH/k$ with respect to the dominant Newtonian terms in~(\ref{Delta_Newt}). On the other hand, the lensing term contains an integral along the line-of-sight that enhances its amplitude at high redshift. 

The other terms in the second and third lines of~(\ref{Delta_pap}) are directly proportional to the Bardeen potentials $\Phi$ and $\Psi$. They are therefore suppressed by $(\HH/k)^2$ with respect to the dominant Newtonian terms in~(\ref{Delta_Newt}), and by $\HH/k$ with respect to the velocity contributions and the gravitational redshift distortions. These terms are purely relativistic, accounting for distortions in the energy of the photons, the length of the geodesic and the time taken by the photons to travel from the galaxy to the observer.

Equation~(\ref{Delta_pap}) does not take into account the fact that, in practice, redshift surveys are magnitude limited: only galaxies above a given luminosity threshold are detected. Since the observed luminosity of galaxies is affected by inhomogeneities along the photons' trajectory (galaxies behind overdense regions are for example magnified), one encounters situations where the intrinsic luminosity of a galaxy is too small to pass the threshold, but the galaxy is detected anyway because it is magnified by inhomogeneities (the opposite also happens: galaxies that should pass the threshold can be demagnified and remain undetected). This effect is called magnification bias and it induces additional fluctuations in the observed overdensities of galaxies of the form~\cite{broadhurst, moessner}
\be
\Delta_{\rm mb}=5s\cdot \kappa\, , \label{mag_bias}
\ee
where $\kappa$ is the convergence and $s$ is the effective number count slope. The standard expression for the convergence contains the effect of gravitational lensing
\be
\kappa=\frac{1}{2r}\int_{0}^{r} dr'\frac{r-r'}{r'}\Delta_\Omega(\Phi+\Psi)\, . \label{kappa_st}
\ee
This term shows how overdense regions situated between the galaxy and the observer magnify the galaxy and change consequently its size and its luminosity. Combined with the first term in the second line of~(\ref{Delta_pap}), equation~(\ref{kappa_st}) gives rise to the standard expression for the magnification bias derived in~\cite{broadhurst, moessner}.

Gravitational lensing is however not the only effect that contributes to the convergence. As shown in~\cite{pap_conv}, $\kappa$ is also affected by relativistic fluctuations. To properly take into account all these contributions, we need to solve Sachs equation~\cite{Sachs} in a perturbed Friedmann universe. Sachs equation tells us how a bundle of geodesics emitted by a distant galaxy is distorted by fluctuations along the photons' trajectory. The fully relativistic expression for the convergence at linear order reads~\cite{pap_conv}
\bea
\fl \kappa=\frac{1}{2r}\int_{0}^{r} dr'\frac{r-r'}{r'}\Delta_\Omega(\Phi+\Psi)+\left(\frac{1}{r\HH} -1\right)\bV\cdot\bn 
-\frac{1}{r}\int_0^r dr' (\Phi+\Psi)\nonumber\\
+\left(1-\frac{1}{r\HH} \right)\int_0^r dr'(\dot{\Phi}+\dot\Psi)+\left(1-\frac{1}{r\HH} \right)\Psi +\Phi \label{kappa_rel} \, .
\eea
The Doppler term in the first line of~(\ref{kappa_rel}) reflects the fact that if a galaxy is moving with respect to the observer, it appears closer or further away in redshift space and consequently its apparent size and luminosity are modified. The sign of this contribution (which determines if the galaxy is magnified or demagnified) depends obviously on the velocity's direction, $\bV\cdot \bn$, but also on the sign of the prefactor 
\be
g(z)\equiv\left(\frac{1}{r\HH} -1\right)\, .
\ee
In a flat $\Lambda$CDM universe with $\Omega_m=0.24$ and $h=0.73$, $g(z)>0$ for $z<1.7$ and  $g(z)<0$ for $z>1.7$. Suppose that we look at a galaxy that is moving toward us, i.e. with $\bV\cdot \bn <0$ ($\bn$ points from the observer to the galaxy). At small redshift, $z<1.7$, the Doppler term demagnifies the galaxy, whereas at high redshift, $z>1.7$, it magnifies it. This can be understood in the following way: for a fixed redshift, a galaxy moving toward us is more distant, in comoving coordinates, than a galaxy with null peculiar velocity. This generates two opposite effects on the observed size of the galaxy. On one hand, a more distant galaxy is observed under a smaller solid angle.
On the other hand, since the galaxy is further away, it has emitted photons at an earlier time, when the scale factor of the universe was smaller. Consequently, the bundle of photons experiences a larger stretch when propagating towards us, which increases the observed size of the galaxy. At small redshift the first effect dominates, leading to a demagnification
of the galaxy, whereas at large redshift the second effect dominates and the galaxy is magnified. At $z \simeq 1.7$,
the two effects compensate, leaving the size of the galaxy unchanged.
The situation is then simply reversed for a source moving away from the observer. 

This velocity term has been called {\it Doppler lensing}~\cite{dopplens} as it describes a change in the magnification due to the Doppler effect. It has been studied in detail in~\cite{pap_conv, dopplens, bacon}. Comparing the Doppler lensing with the standard gravitational lensing, we see that the two terms are in principle of the same order of magnitude $V\sim k\Psi \sim \Delta_\Omega(\Phi+\Psi)$. At small redshift however, the Doppler lensing is enhanced by the factor $1/(\HH r)$ and it consequently dominates over the gravitational lensing term. At large redshift on the other hand, the gravitational deflections along the line-of-sight accumulate and the gravitational lensing dominates over the Doppler lensing. Note that the first part of the Doppler lensing term, proportional to $1/(\HH r)$, is included in the Newtonian expression~(\ref{Delta_Kaiser}) where the variation in the luminosity function is related to the slope $s$ by
\be
\frac{d\ln \phi}{d\ln r}=-5 s\, .
\ee 

The other terms in the convergence~(\ref{kappa_rel}) are suppressed by a factor $\HH/k$ with respect to the gravitational lensing and the Doppler lensing. These terms represent relativistic corrections to the observed distance (and consequently the apparent size) of the galaxies generated by the gravitational potentials $\Phi$ and $\Psi$. 

Combining equations~(\ref{Delta_pap}), (\ref{mag_bias}) and (\ref{kappa_rel}) we obtain the full expression for $\Delta$, valid in any metric theory of gravity, in the linear regime.

\section{Correlation functions}

Let us now study the correlation functions of $\Delta$. As seen above, the relativistic contributions to $\Delta$ are suppressed by powers of $\HH/k$ with respect to the dominant Newtonian terms of equation~(\ref{Delta_Newt}). The relativistic effects are therefore expected to leave a detectable imprint on the correlation functions at large scales mainly, when $k\sim \HH$. To study consistently this regime, we need statistical measures that do not intrinsically rely on small-scale approximations, like for example the distant-observer approximation. Here we review two statistical measures that are well adapted to the large-scale regime: the angular power spectrum, $C_\ell(z,z')$, and the two-point correlation function in configuration space, $\xi(z, z', \theta)$. Note that the spherical Fourier Bessel power spectrum, $C_\ell(k,k')$, is also well suited to the full-sky regime and provides an interesting alternative~\cite{sfb}.    

\subsection{Angular power spectrum}

\label{sec:Cl}

At fixed redshift, $\Delta(z,\bn)$ is a function on the sphere and we can naturally expand it in spherical harmonics
\be
\Delta(z, \bn) =\sum_{\ell m}a_{\ell m}(z)Y_{\ell m}(\bn) \, .
\ee
The angular power spectrum is defined as
\be
C_\ell (z, z') = \langle a_{\ell m}(z)a^*_{\ell m}(z') \rangle \, , \label{Cl}
\ee
where a star indicates complex conjugation.
We use the Fourier convention
\be
f(z, \bx)=\frac{1}{(2\pi)^3}\int d^3\bk\, e^{-i\bk\bx}f(z, \bk)\, ,
\ee
and we relate the density, velocity and Bardeen's potentials to the initial metric perturbation $\Psi_{\rm in}(\bk)$ via the transfer functions
\bea
D(z, \bk)&=&T_D(z, k)\Psi_{\rm in}(\bk)\, , \label{TD}\\
V(z, \bk)&=&T_V(z, k)\Psi_{\rm in}(\bk)\, ,\\
\Psi(z, \bk)&=&T_\Psi(z, k)\Psi_{\rm in}(\bk)\, ,\\
\Phi(z, \bk)&=&T_\Phi(z, k)\Psi_{\rm in}(\bk)\, . \label{TPhi}
\eea
The initial power spectrum is characterized by the amplitude $A$ and the scalar spectral index $n_s$
\be
k^3\langle\Psi_{\rm in}(\bk)\Psi_{\rm in}(\bk') \rangle=(2\pi)^3 A (k\eta_0)^{n_s-1} \delta(\bk+\bk')\, . \label{Pini}
\ee 
The time and $k$-dependence of the transfer functions (\ref{TD}) to (\ref{TPhi}) depend on the cosmology. In particular the growth rate of perturbations is sensitive to the nature of dark energy and to the theory of gravitation.
In general relativity, one has
\bea
T_\Psi(z, k)=\frac{D_1(a)}{a}T(k)\, ,\label{TPsi_GR}\\
T_\Phi(z, k)=T_\Psi(z, k)\, ,\\
T_D(z, k)=-\frac{2D_1(a)}{3\Omega_m}\left(\frac{k}{\HH_0}\right)^2 T(k)\, ,\\
T_V(z, k)=\frac{2a}{3\Omega_m}\frac{ \HH}{\HH_0}\frac{k}{\HH_0}\left(T_\Psi+\frac{1}{\HH}\dot{T}_\Phi \right)=
\frac{2}{3\Omega_m}\frac{\HH}{\HH_0}\frac{k}{\HH_0}f(a)D_1(a)T(k)\, , \label{TV_GR}
\eea
where $T(k)$ is a time-independent transfer function, $D_1$ denotes the linear growth function and $f$ is the growth rate defined through
\be
f(a)=\frac{d\ln D_1(a)}{d\ln a}\, .
\ee
Inserting equations~(\ref{TD}) to~(\ref{Pini}) into (\ref{Cl}), the angular power spectrum can be expressed as~\cite{pap_gal}
\be
C_\ell(z,z')=\frac{2A}{\pi}\int \frac{dk}{k}(k\eta_O)^{n_s-1}\Delta_\ell(z, k)\Delta_\ell(z', k)\, , \label{Clint}
\ee
with
\bea
\fl\Delta_\ell(z, k)= b\, T_D j_\ell(kr)+\frac{k}{\HH}T_Vj''_\ell(kr) +\ell(\ell+1)(2-5s)\int_0^r dr'\frac{r-r'}{2rr'}\big(T_\Phi+T_\Psi \big)j_\ell(kr')\nonumber\\
\fl+\left[\left(\frac{\dot\HH}{\HH^2} +5s-1 +\frac{2-5s}{r\HH}\right)T_V-\frac{1}{\HH}\dot T_V +\frac{k}{\HH}T_\Psi \right] j'_\ell(kr)\nonumber\\
\fl +\frac{2-5s}{r}\int_0^r dr' \left(T_\Phi+T_\Psi \right)j_\ell(kr')
 +\left (\frac{\dot\HH}{\HH^2} +5s +\frac{2-5s}{r\HH}\right)\int_0^r dr' \left(\dot T_\Phi+\dot T_\Psi \right)j_\ell(kr')\nonumber\\
\fl +\left[\left (\frac{\dot\HH}{\HH^2} +5s+1 +\frac{2-5s}{r\HH}\right)T_\Psi +
(5s-2)T_\Phi +\frac{1}{\HH}\dot T_\Phi -3\frac{\HH}{k}T_V \right] j_\ell(kr)\, .\label{Deltal}
\eea
Here $j_\ell$ denotes the spherical Bessel function of order $\ell$ and a prime is a derivative with respect to its argument $kr$. Expression~(\ref{Deltal}) agrees with equation (2.17) of~\cite{montanari_class}, in the case where Euler equation~(\ref{Euler}) is valid. 

The angular power spectrum~(\ref{Clint}) is a function of $z$, $z'$ and $\ell$. In practice, galaxy surveys are split into redshift bins, and the $C_\ell$'s are measured either within the same bin or as cross-correlation between two different bins~\cite{pap_gal, montanari_cl}
\be
\label{Clwin}
\fl\hat C_\ell (z,z')=\frac{2A}{\pi}\int \frac{dk}{k}(k\eta_O)^{n_s-1}\int_0^\infty\! d\hat z\, W(z,\hat z)\Delta_\ell(k,\hat z)
\int_0^\infty \!d\hat z'\, W(z',\hat z')\Delta_\ell(k, \hat z')\, ,
\ee
where the window function $W(z,\hat z)$ determines the size and shape of the redshift bin centred at $z$. For example, for a Gaussian window function of width $\sigma_z$
\be
W(z,\hat z)=\frac{1}{\sqrt{2\pi}\sigma_z}\exp\Big[-(z-\hat z)^2/(2\sigma_z^2)\Big]\, .
\ee

\begin{figure}[!t]
\centerline{\epsfig{figure=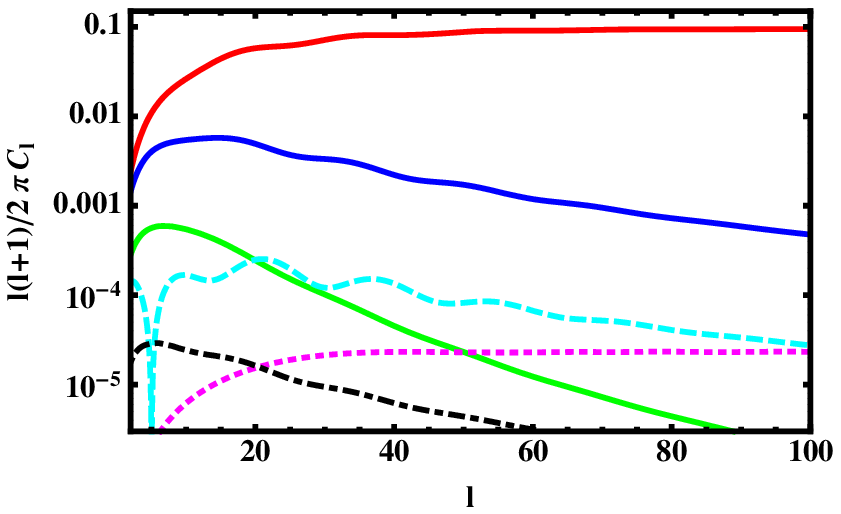,width=0.5\textwidth}\hspace{0.3cm}\epsfig{figure=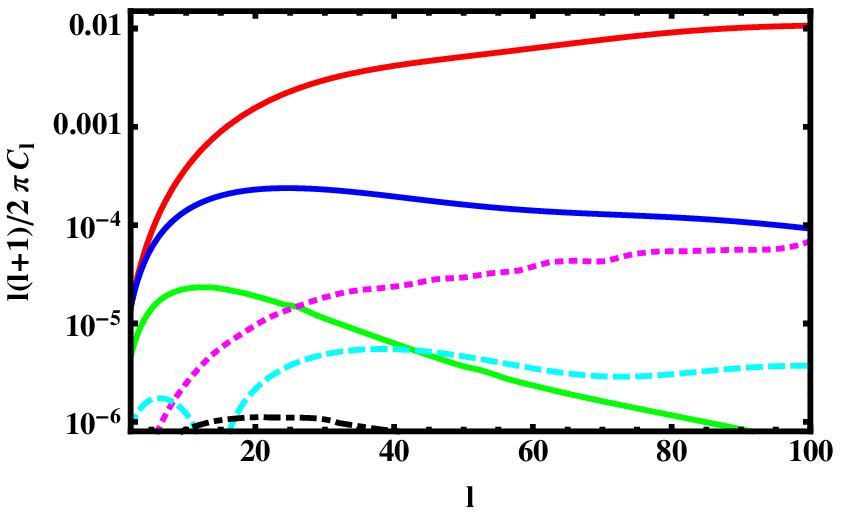,width=0.5\textwidth}}
\centerline{\epsfig{figure=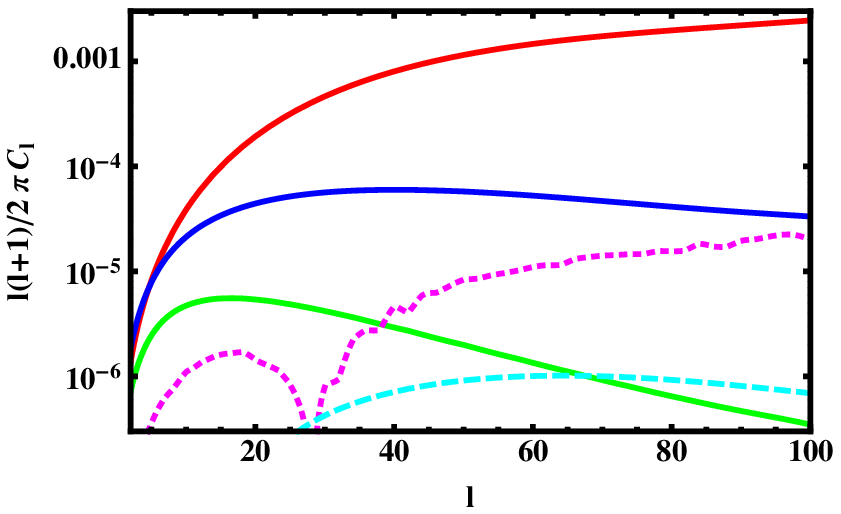,width=0.5\textwidth}\hspace{0.3cm}\epsfig{figure=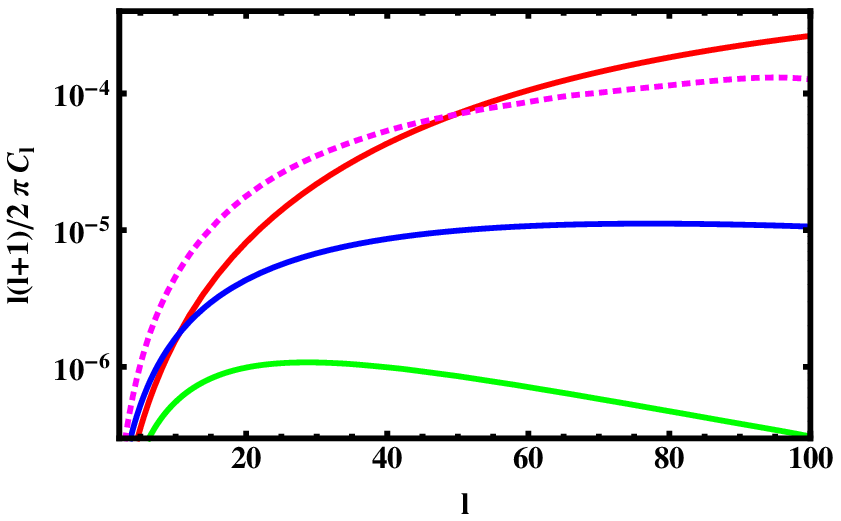,width=0.498\textwidth}}
\caption{ \label{fig:cls} The angular power spectrum, binned with a Gaussian window function of width $\sigma_z=0.1\,z$, plotted as a function of $\ell$, for redshift bins centred at $z=0.1$ (top left), $z=0.5$ (top right), $z=1$ (bottom left) and $z=3$ (bottom right). The different curves denote the contribution from: density (red, top solid line), redshift-space distortions (green, bottom solid line), the correlation of density with redshift-space distortions (blue, middle solid line), lensing (magenta, dotted line), Doppler (cyan, dashed line) and gravitational potential (black, dashed-dotted line). We have chosen $b=1$ and $s=0$ (no magnification bias).}
\end{figure}

In figure~\ref{fig:cls} we plot the angular power spectrum as a function of $\ell$, calculated with a Gaussian window function of width $\sigma_z=0.1\, z$, for fixed redshift bins $z=z'$. The calculation is done in a flat $\Lambda$CDM universe with $\Omega_m=0.24$, $h=0.73$, $\sigma_8=0.75$ and $n_s=1$. We use the transfer functions (\ref{TPsi_GR}) to (\ref{TV_GR}) and $T(k)$ is calculated with CAMB~\cite{camb}.
The different curves represent different contributions to the $C_\ell$'s: the density (red, top solid line) refers to the first term in equation~(\ref{Deltal}), redshift distortion (green, bottom solid line) refers to the second term, lensing (magenta, dotted line) to the third term, Doppler (cyan, dashed line) to the second line (we have used here Euler equation to relate the gravitational redshift distortions to the peculiar velocity), and gravitational potential (black, dashed-dotted line) refers to the third and fourth lines of~(\ref{Deltal}). Each contribution includes its auto-correlation as well as its cross-correlation with density and redshift distortions. The only exception is the cross-correlation between the density and redshift distortions which is explicitly shown in blue (solid middle line). Note that the cross-correlations between the lensing, Doppler and potential terms are completely subdominant and are therefore not included in figure~\ref{fig:cls}. 

At small redshift, we see that the dominant correction to the standard redshift distortions expression~(\ref{Delta_Newt}) is due to the Doppler term (cyan, dashed line). At large redshift however, the main correction comes from the gravitational lensing term (magenta, dotted line), which becomes larger than the density term (red, top solid line) at $z=3$. Future galaxy surveys, that will be able to measure the angular power spectrum up to large separation (small $\ell$) and high redshift, will have to consistently take into account these two types of contributions. The gravitational potential corrections (black, dashed-dotted line) on the other hand remain small at all scales and redshifts and can be neglected.

There are two main advantages of working with the angular power spectrum. First, measuring the $C_\ell(z, z')$'s does not require any knowledge of the cosmology. The redshifts $z$ and $z'$ and the directions $\bn$ and $\bn'$ can be used directly to express the angular power spectrum as a function of $z, z'$ and $\ell$. Theses measurements can then be confronted with theoretical predictions, allowing to probe without ambiguity the transfer functions. This is an advantage with respect to observations in Fourier space, where the redshifts $z$ and $z'$ need to be translated into the comoving distances $r$ and $r'$, in order to calculate the Fourier transforms. Since the transformation from redshifts to comoving coordinates depends on the background cosmology, such a procedure is vulnerable to errors in this translation due to our imperfect knowledge of the cosmological parameters. The $C_\ell$'s, that do not suffer from this problem, are optimally adapted to the coordinate system in which measurements are made $(z, z', \bn\cdot \bn')$.

The second advantage of the angular power spectrum is that it is extremely well suited to measure correlations at large separations. The expansion in spherical harmonics naturally includes wide-angle effects. The expressions~(\ref{Clint}) and (\ref{Deltal}) are valid in the full-sky, and can easily be calculated without requiring any flat-sky approximation. The $C_\ell$'s are therefore particularly well suited to probe the relativistic effects, since they become important at large separations.

There are some disadvantages however in using the angular power spectrum. First, the $C_\ell$'s project $\Delta$ onto two-dimensional spheres at fixed redshift. As shown in~\cite{cl_lewis}, no information is lost through this projection, as long as one takes into account sufficiently thin redshift bins and measures not only auto-correlations but also cross-correlations between bins. However, this projection has the disadvantage of mixing the different terms in $\Delta$. This is not the case in Fourier space and in configuration space, where for example the density and the redshift-space distortions can be separated by using a multipole expansion in the angle $\beta$~\cite{Kaiser_rsd, hamilton} (here $\beta$ denotes the orientation of the galaxy pair with respect to the observer, see figure~\ref{fig:cone}). This splitting has no equivalent in the angular power spectrum, since a given $\ell$ mixes different angles $\beta$. 

\begin{figure}[t]
\centerline{\epsfig{figure=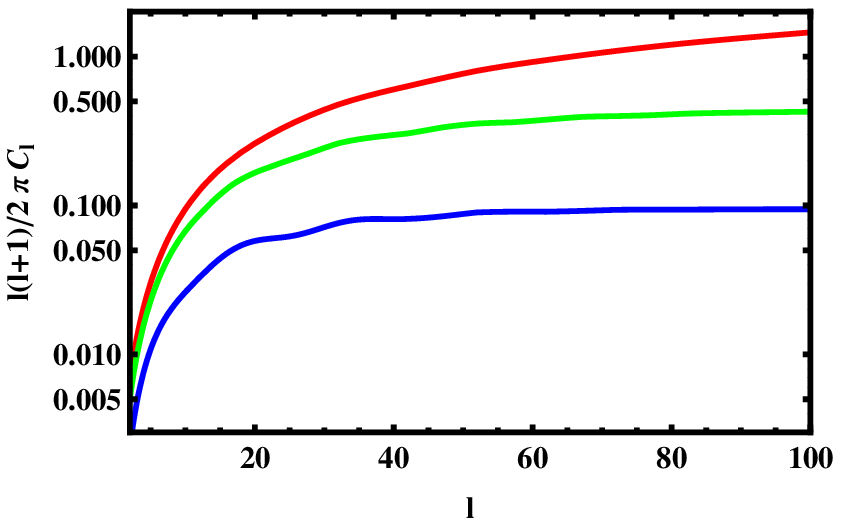, width=0.49\textwidth}\hspace{0.3cm}\epsfig{figure=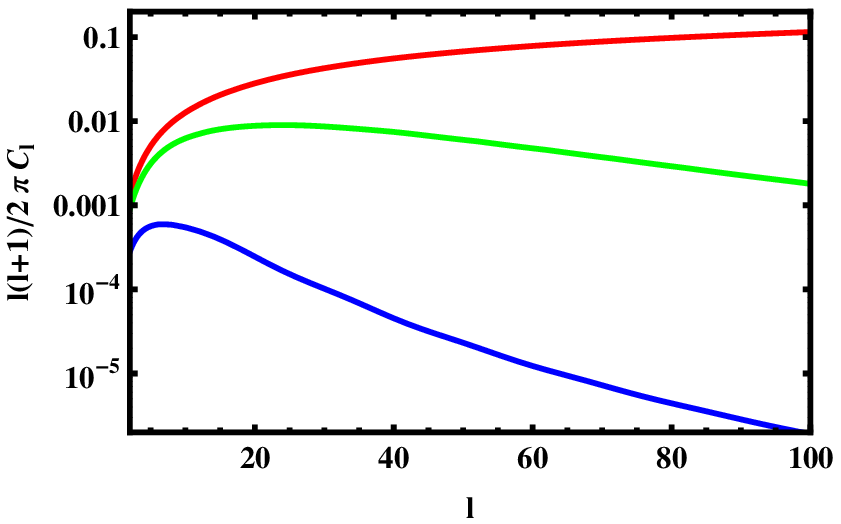, width=0.49\textwidth}}
\caption{ \label{fig:window} The angular power spectrum at $z=z'=0.1$ for the density contribution (left panel) and the redshift-space distortions (right panel) with different redshift binnings. The top curves (red) have no binning, the middle curves (green) have a Gaussian window function with $\sigma_z=0.002$ and the bottom curves (blue) have a Gaussian window function with $\sigma_z=0.01$.}
\end{figure}

The second inconvenient of the $C_\ell$'s is to be quite sensitive to the redshift binning, even at large separation. Figure~\ref{fig:window} shows the density (left panel) and the redshift distortions (right panel) calculated with different sizes of redshift bins. The amplitude of the angular power spectrum strongly decreases with the size of the redshift bin. This strong dependence can be understood by looking at the expressions for the angular power spectrum. Without any binning in redshifts, we see from equations~(\ref{Clint}) and~(\ref{Deltal}) that when $z=z'$,  the $C_\ell$'s are sensitive to $j^2_\ell(kr)$ which is always positive. The different wavenumbers add therefore in a constructive way in the integral~(\ref{Clint}).  On the other hand, when a window function is introduced~(\ref{Clwin}), the $\hat C_\ell$'s are sensitive to $j_\ell(kr(\hat z))j_\ell(kr(\hat z'))$ with $\hat z\neq \hat z'$. The spherical Bessel functions are not in phase anymore and contribute sometimes positively, sometimes negatively to the integral over $k$, decreasing the overall amplitude of the angular power spectrum. Since the redshift binning does not affect all contributions in the same way (for example redshift distortions decrease faster than the density with the size of $\sigma_z$), it is important to know quite well the binning used in galaxy surveys in order not to misinterpret the amplitude of each contribution to the signal. 

\subsection{Two-point correlation function}

\begin{figure}[t]
\centerline{\epsfig{figure=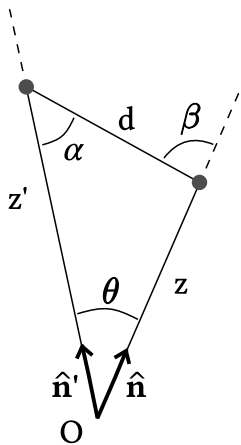,height=5.8cm}}
\caption{ \label{fig:cone} The triangle formed by the observer O and the two galaxies at positions $(z,\bn)$ and $(z',\bn')$. The variable $d$ represent the comoving distance between the two galaxies. The angle $\theta$, $\beta$ and $\alpha$ are also shown.}
\end{figure}

An alternative statistical measure to the angular power spectrum is the two-point correlation function
\be
\xi=\langle \Delta(z,\bn) \Delta(z',\bn')\rangle\, .
\ee
The function $\xi$ depends on the shape of the triangle formed by the observer and the two galaxies (see figure~\ref{fig:cone}). Due to the statistical isotropy of our universe, $\xi$ can be expressed as a function of three quantities only, for example $(z, z', \theta)$ or $(z, d, \beta)$, where $\theta$ denotes the angle between $\bn$ and $\bn'$, and $\beta$ denotes the angle between $\bn$ and the galaxies' separation $\bN$. The coordinate system $(z, z',\theta)$ has the advantage of being the system in which measurements are performed. As the angular power spectrum, $\xi(z, z',\theta)$ is therefore directly observable and does not rely on any cosmological assumption. Note that the function $\xi(z, z', \theta)$ probes roughly the same geometrical configurations as $C_\ell(z, z')$, with $\theta\sim \pi/\ell$. 

The coordinate system $(z, d, \beta)$ on the other hand requires to translate the redshifts and the separation angle into the comoving distance $d$ and the angle $\beta$. This translation has the inconvenient to rely on a choice for the background cosmological parameters. However, this coordinate system has the strong advantage of exploiting in a natural way the symmetries of the different contributions in $\Delta$. It is therefore the optimal system to separate the density from the redshift distortions and the relativistic contributions. 

In this coordinate system, the density term is extremely simple since it is independent of $\beta$.
Indeed, due to the statistical homogeneity and isotropy of the primordial density fluctuations, the correlation
\be
\xi^{\rm dens}=\langle b(z)D(z,\bn) b(z')D(z',\bn')\rangle
\ee
depends only on the comoving pair separation $d$ and on the redshift $z$.

Redshift-space distortions on the other hand depend explicitly on $\beta$. The correlation function
\be
\fl\xi^{\rm rsd}=\left\langle \left(b(z) D(z,\bn)-\frac{1}{\mathcal{H}(z)}\partial_r(\mathbf{V}\cdot\mathbf{n})\right)
\left( b(z')D(z',\bn')-\frac{1}{\mathcal{H}(z')}\partial_{r'}(\mathbf{V}\cdot\mathbf{n'})\right)\right\rangle
\ee   
is sensitive to the orientation of the pair of galaxies with respect to the observer. Hamilton showed in~\cite{hamilton} that, in the distant-observer approximation, the dependence of $\xi^{\rm rsd}$ in $\beta$ is very simple. The cross-correlation between density and velocities generates a quadrupole modulation, proportional to $P_2(\cos\beta)$, whereas the velocity-velocity correlation generates an hexadecapole, proportional to $P_4(\cos\beta)$. 
By fitting for a monopole, a quadrupole and an hexadecapole in the correlation function, one can therefore separate the density contribution from the velocity contribution. The monopole and quadrupole have been measured in various galaxy surveys (see e.g. the latest measurements from BOSS~\cite{boss_rsd}), and they have been used to probe the validity of general relativity through measurements of the growth rate $f$.

The multipole expansion in the angle $\beta$ can be extended to the relativistic contributions in $\Delta$. The Doppler contribution and the gravitational redshift distortions
\be
\Delta^{\rm rel\, V}=\frac{1}{\HH}\dd_r\Psi + \frac{1}{\HH}\dot{\bV}\cdot\bn\label{Drelgen}
-\left[\frac{\dot{\HH}}{\HH^2}+5s-1+\frac{2-5s}{r\HH}\right]\bV\cdot\bn
\ee
are intrinsically different from the density and redshift distortions since they break the symmetry of the correlation function, i.e. they generate odd multipoles in $\beta$, namely a dipole, proportional to $P_1(\cos\beta)$, and an octupole, proportional to $P_3(\cos\beta)$~\cite{pap_asymm, roy}. The derivation of these new multipoles is shown below.

These odd multipoles cannot be measured in the standard way in galaxy surveys since by construction $\langle\Delta(z,\bn)\Delta(z',\bn') \rangle$ is symmetric under the exchange of $(z,\bn)$ and $(z',\bn')$. The standard procedure to extract a multipole of order $\ell$ consists in weighting each pair of galaxies by the appropriate Legendre polynomial $P_\ell(\cos\beta)$ and then integrating over all values of $\beta$
\be
\xi_\ell(z,d)=\frac{2\ell+1}{2}\int_{-1}^1 d\mu\, \xi(z, d, \beta)\,P_\ell(\mu)\, , \quad {\rm with} \quad \mu=\cos\beta\, . \label{mult}
\ee
In practice, each pair is counted twice, the first time $\Delta(z,\bn)\Delta(z',\bn')$ is weighted by $P_\ell(\cos\beta)$ and the second time $\Delta(z',\bn')\Delta(z,\bn)$ is weighted by $P_\ell(\cos\beta')$ where $\beta'=\beta+\pi$. Due to the symmetry of $\langle\Delta(z,\bn)\Delta(z',\bn') \rangle$, the two contributions add up if $\ell$ is even and cancel out if $\ell$ is odd (since for odd $\ell$, $P_\ell(\cos(\beta+\pi))=-P_\ell(\cos\beta)$). 

The breaking of symmetry generated by the Doppler and the gravitational redshift terms is therefore automatically washed out in $\langle\Delta(z,\bn)\Delta(z',\bn') \rangle$. On the other hand if one looks at cross-correlations between two populations of galaxies with different $\Delta$, like for example a bright population (denoted by B) and a faint population (denoted by F), then $\langle\Delta_{\rm B}(z,\bn)\Delta_{\rm F}(z',\bn') \rangle$ is not intrinsically symmetric anymore under the exchange of $(z,\bn)$ and $(z',\bn')$. Due to the Doppler and the gravitational redshift terms, $\langle\Delta_{\rm B}(z,\bn)\Delta_{\rm F}(z',\bn') \rangle$ now differs from $\langle\Delta_{\rm F}(z,\bn)\Delta_{\rm B}(z',\bn') \rangle$ and this difference generates odd multipoles in the correlation function (\ref{mult}). Note that this breaking of symmetry has been discussed for the first time by McDonalds~\cite{mcdonalds} and Yoo et al.~\cite{uros} in Fourier space and by Croft~\cite{croft} in configuration space.

\begin{figure}[!t]
\centerline{\epsfig{figure=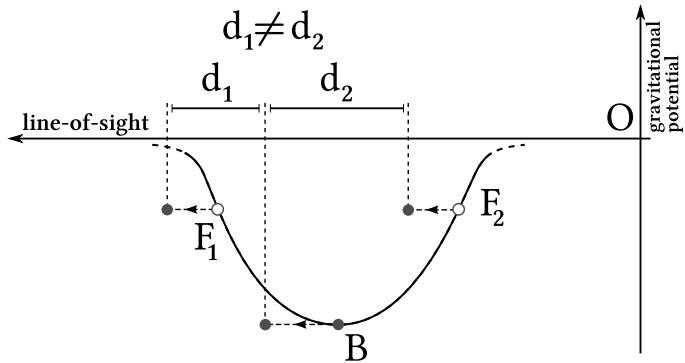,height=4.8cm}}
\caption{ \label{fig:grav_red} Sketch of the gravitational redshift
  effect. The observer is sitting at O. The bright galaxy B suffers the largest
gravitational redshift because it is sitting at the bottom of the
potential well. The faint galaxies $\F_1$ and $\F_2$ shift by a somewhat smaller amount.
As a consequence the distances $d_1\neq d_2$. 
This generates an asymmetric cross-correlation
function: the bright galaxy is differently correlated with faint galaxies behind it
than in front of it.}
\end{figure}

The physical origin of the symmetry breaking can be understood by looking at figure~\ref{fig:grav_red}. The observer, sitting at O, detects three galaxies along its line-of-sight, one bright galaxy B, and two faint galaxies ${\rm F}_1$ and ${\rm F}_2$ symmetrically situated around B. Let us assume that the bright galaxy is at the bottom of a gravitational potential. Due to gravitational redshift, the bright galaxy appears redshifted with respect to the observer. Gravitational redshift also shifts the positions of the two faint galaxies ${\rm F}_1$ and ${\rm F}_2$, but by a smaller amount than the bright one (simply because they are higher up in the gravitational potential). As a consequence, in redshift-space the situation is not symmetric anymore: the faint galaxy behind, ${\rm F}_1$, is closer to the bright galaxy, than the faint galaxy in front, ${\rm F}_2$. The correlation function between B and ${\rm F}_1$ differs therefore from the correlation function between B and ${\rm F}_2$. This situation is highly idealised, but the principle is sound: gravitational redshift statistically breaks the symmetry of the correlation function. One can show that the Doppler terms in~(\ref{Drelgen}) induce a similar breaking of symmetry. This is not surprising since the velocity, $\bV\cdot\bn$, also contains one single gradient of the gravitational potential.  

\subsubsection{Multipole expansion} 

Let us now explicitly calculate the multipole expansion of the correlation function. 
As shown in~\cite{szalay, szapudi, szapudi2, francesco, bertacca}, the function $\xi$ can naturally be expanded in terms of sinus and cosinus of the angles $\beta$ and $\alpha$ defined in figure~\ref{fig:cone}. Note that we unambiguously define the angle $\beta$ with respect to the bright galaxy in the pair, which we place at position $(z,\bn)$.

The full expression $\langle \Delta_{\rm B}(z, \bn)\Delta_{\rm F}(z', \bn)\rangle$ contains a large number of terms. Here we concentrate on the three dominant contributions, namely the standard redshift-space distortions correlation 
\be
\xi^{\rm rsd}=\langle \Delta^{\rm rsd}_{\rm B}(z, \bn)\Delta^{\rm rsd}_{\rm F}(z', \bn)\rangle\, ,
\ee
the cross-correlation between the standard redshift distortions and the dominant relativistic corrections
\be
\xi^{\rm rel}=\langle \Delta^{\rm rsd}_{\rm B}(z, \bn)\Delta^{\rm rel\, V}_{\rm F}(z', \bn)\rangle+ \langle \Delta^{\rm rel\, V}_{\rm B}(z, \bn)\Delta^{\rm rsd}_{\rm F}(z', \bn)\rangle\, ,
\ee
and the cross-correlation between the standard redshift distortions and the gravitational lensing
\be
\xi^{\rm lens}=\langle \Delta^{\rm rsd}_{\rm B}(z, \bn)\Delta^{\rm lens}_{\rm F}(z', \bn)\rangle+ \langle \Delta^{\rm lens}_{\rm B}(z, \bn)\Delta^{\rm rsd}_{\rm F}(z', \bn)\rangle\, . \label{xilens}
\ee
Here we have defined, for the bright population, the standard redshift distortions as
\be
\Delta^{\rm rsd}_{\rm B}=b_{\rm B} D-\frac{1}{\mathcal{H}}\partial_r(\mathbf{V}\cdot\bn)\, ,
\ee
the dominant relativistic corrections as
\be
\Delta^{\rm rel\, V}_{\rm B}=\frac{1}{\HH}\dd_r\Psi + \frac{1}{\HH}\dot{\bV}\cdot\bn
-\left[\frac{\dot{\HH}}{\HH^2}+5s_{\rm B}-1+\frac{2-5s_{\rm B}}{r\HH}\right]\bV\cdot\bn
\ee
and the lensing term as
\be
\Delta^{\rm lens}_{\rm B}=(5s_{\rm B}-2)\int_0^r dr'\frac{(r-r')r'}{2 r}\nabla_\perp^2 (\Phi+\Psi)\, .
\ee
The faint population has similar expressions, with the bias $b_{\rm B}$ replaced by $b_{\rm F}$ and the slope $s_{\rm B}$ replaced by $s_{\rm F}$. One can show~\cite{pap_asymm} that the auto-correlation of the relativistic term $\langle \Delta^{\rm rel\, V}_{\rm B}\Delta^{\rm rel\, V}_{\rm F} \rangle$ and the auto-correlation of the lensing terms $\langle \Delta^{\rm lens}_{\rm B}(\bn, z)\Delta^{\rm lens}_{\rm F} (\bn, z')\rangle$ are much smaller than their cross-correlation with the standard redshift distortions. Therefore we do not include them here.

Using the tripolar expansion defined in~\cite{szapudi, szapudi2}, we find for the standard redshift-space correlation function
\be
\label{xistalpha}
\fl \xi^{\rm rsd}=\frac{2A}{9\pi^2\Omega_m^2}\Big\{S_1+S_2\cos(2\beta)+S_3\cos(2\alpha)
+S_4\cos(2\alpha)\cos(2\beta)
+S_5\sin(2\alpha)\sin(2\beta) \Big\}\, ,
\ee
where the coefficients $S_1$ to $S_5$ are given in~\ref{app:coeffst}. They depend on $r, r'$ and $d$.
The relativistic correlation function can be calculated using the same method~\cite{francesco, bertacca}. We find
\bea
\label{xirelalpha}
\fl \xi^{\rm rel}=\frac{2A}{9\pi^2\Omega_m^2}\Bigg\{R_1 \cos(\alpha)+R_2\cos(\beta)
+R_3\cos(\alpha)\cos(2\beta)\\+R_4\cos(\beta)\cos(2\alpha)
+R_5\sin(\alpha)\sin(2\beta)+R_6\sin(\beta)\sin(2\alpha)\Bigg\}\, ,\nonumber
\eea
where the coefficients $R_1$ to $R_6$ are given in~\ref{app:coeffrel}. Note that here the coefficients $S_i$ and $R_i$ have been calculated assuming general relativity, i.e. using equations (\ref{TPsi_GR}) to (\ref{TV_GR}) and using Euler equation~(\ref{Euler}).

Comparing equation (\ref{xistalpha}) with equation (\ref{xirelalpha}), we see immediately that the symmetries of the two expressions are different. The standard redshift-space correlation function contains only contributions sensitive to $2\beta$ or $2\alpha$, that are invariant under a change $\beta\rightarrow \beta+\pi$ and $\alpha\rightarrow \alpha+\pi$, i.e. under the exchange of the relative position of the bright and the faint galaxy~\footnote{Note that at large separation, the exchange of the positions $(z,\bn)$ and $(z',\bn')$ results in $\beta\rightarrow \beta+\pi-\theta$, as we will discuss below.}. The relativistic correlation function on the other hand contains terms that are directly sensitive to $\beta$ and $\alpha$ and that consequently change sign when $\beta\rightarrow \beta+\pi$ or $\alpha\rightarrow \alpha+\pi$. 

We can now calculate the multipoles of the correlation function using equation~(\ref{mult}). We first need to express $\alpha$ and $r'$ in terms of $\beta, \, r$ and $d$ using
\bea
r'=\sqrt{r^2+d^2+2dr\cos(\beta)}\label{rp}\, ,\\
\cos(\alpha)=\frac{d+r\cos(\beta)}{\sqrt{r^2+d^2+2dr\cos(\beta)}}\label{cos}\, ,\\
\sin(\alpha)=\frac{r\sin(\beta)}{\sqrt{r^2+d^2+2dr\cos(\beta)}}\label{sin}\, ,
\eea
and then we can numerically integrate $\xi$ over $\beta$ to extract each multipole. This method is valid at all scales. 

\subsubsection{Distant-observer approximation}

To gain knowledge about the symmetries of the standard and the relativistic correlation functions, it is useful to simplify equations (\ref{xistalpha}) and (\ref{xirelalpha}) using the distant-observer approximation. 
We know that in this approximation, the standard redshift-space correlation function reduces to the sum of a monopole, a quadrupole and an hexadecapole. We want to find an analogous simple expression for the relativistic correlation function. 
The distant-observer approximation is valid when $d\ll r$. In this case, we can expand equations (\ref{rp}) to (\ref{sin}) in series of $d/r$. At lowest order, we find for the standard redshift-space correlation function
\bea
\label{xistcos}
\fl \xi^{\rm rsd}(r, d, \beta)= \frac{2AD_1^2}{9\pi^2\Omega_m^2}
\Bigg\{\Bigg[b_{\rm B}b_{\rm F}+(b_{\rm B}+b_{\rm F})\frac{f}{3}+ \frac{f^2}{5}\Bigg]\mu_0(d)\\
-\Bigg[(b_{\rm B}+b_{\rm F})\frac{2f}{3} +\frac{4f^2}{7}  \Bigg]\mu_2(d) \cdot P_2(\cos\beta)
+\frac{8f^2}{35}\mu_4(d)\cdot P_4(\cos\beta)\Bigg\}\, , \nonumber
\eea
where the functions $D_1, f$ and $b$ are evaluated at $r$ and
\be
\mu_\ell(d)=\int\frac{dk}{k}\left(\frac{k}{\HH_0}\right)^4(k\eta_0)^{n_s-1}T^2(k)j_\ell(kd)\,, \hspace{0.1cm} \ell=0,2,4\,. \label{mul}
\ee
For a single population of galaxies, with $b_{\rm B}=b_{\rm F}$, we recover the well-known Hamilton expression~\cite{hamilton}.

The relativistic correlation function can be simplified in the same way. At lowest order in $d/r$, it becomes~\cite{pap_asymm}
\bea
\label{xirelcos}
\fl \xi^{\rm rel}(r,  d, \beta)=\frac{2A}{9\pi^2\Omega_m^2}\frac{\HH}{\HH_0} D_1^2 f \Bigg\{
\Bigg[\left(\frac{\dot{\HH}}{\HH^2}+\frac{2}{r\HH}\right)(b_{\rm B}-b_{\rm F})\\
+\left(\frac{1}{r\HH}-1\right)\Big(5\big(s_{\rm F}b_{\rm B}-s_{\rm B}b_{\rm F}\big)
-3(s_{\rm B}-s_{\rm F})f \Big)\Bigg]\nu_1(d) \cdot P_1(\cos\beta)\nonumber \\
+ 2\left(\frac{1}{r\HH}-1\right)(s_{\rm B}-s_{\rm F})f \cdot\nu_3(d) \cdot P_3(\cos\beta)\Bigg\}  \, ,\nonumber
\eea
with
\be
\nu_\ell(d)=\int\frac{dk}{k}\left(\frac{k}{\HH_0}\right)^3(k\eta_0)^{n_s-1}T^2(k)j_\ell(kd)\,, \hspace{0.1cm} \ell=1,3 \, . \label{nul}
\ee
We see that the relativistic correlation function contains a dipole, proportional to $P_1(\cos\beta)$, and an octupole, proportional to $P_3(\cos\beta)$, that are not present in the Hamilton expression. These odd multipoles are a direct signature of the relativistic corrections in $\Delta$. By fitting for a dipole and an octupole, besides the standard monopole, quadrupole and hexadecapole, we can consequently target the relativistic effects and extract new information from $\Delta$. 

\begin{figure}[!t]
\centerline{\epsfig{figure=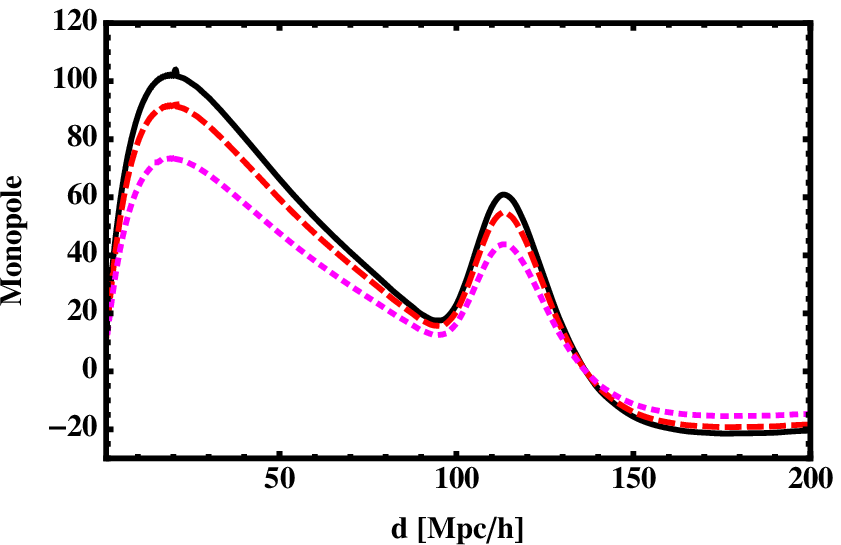,height=5.1cm}\hspace{0.1cm}\epsfig{figure=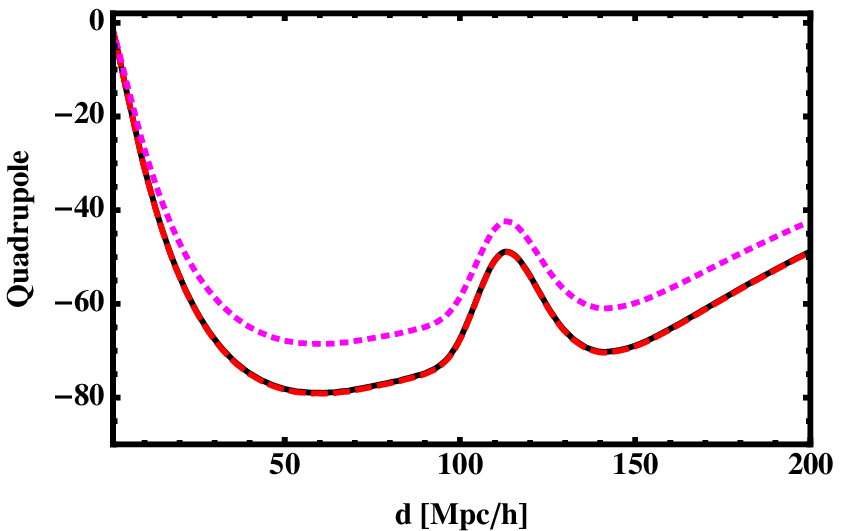,height=5.1cm}}
\centerline{\epsfig{figure=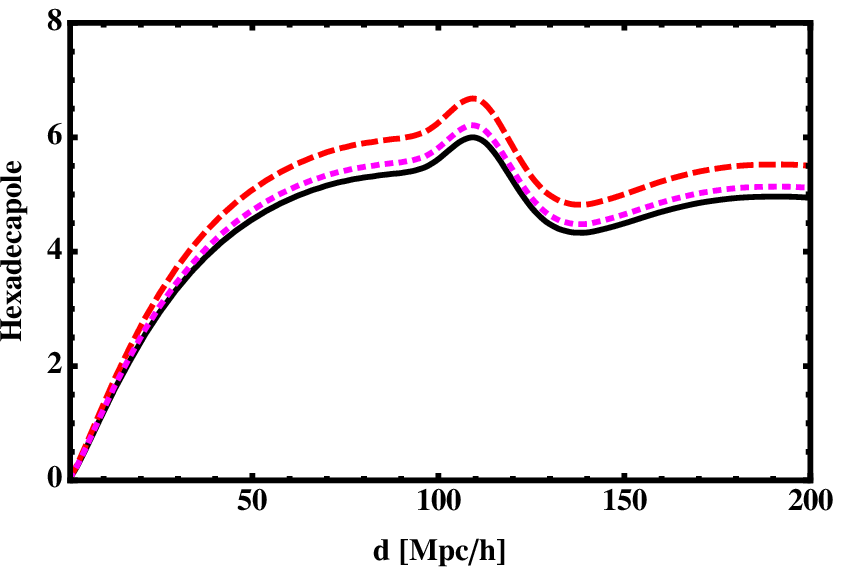,height=5.1cm}\hspace{0.3cm}\epsfig{figure=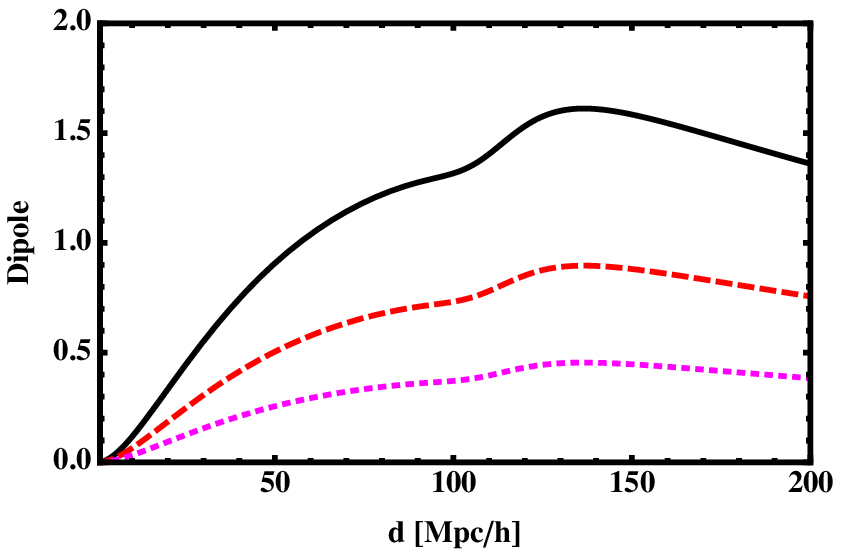,height=5.1cm}}
\caption{ \label{fig:multipole} Amplitude of the monopole (top left), quadrupole (top right), hexadecapole (bottom left) and dipole (bottom right), as a function of the separation $d$. The multipoles are multiplied by $d^2$. In each plot, the black solid line is at $z=0.25$, the red dashed line at $z=0.5$ and the magenta dotted line at $z=1$.}
\end{figure}

In figure~\ref{fig:multipole}, we plot the amplitude of the different multipoles as a function of the separation $d$ at three different redshifts $z=0.25$ (black solid line), $z=0.5$ (red dashed line) and $z=1$ (magenta dotted line). We choose a $\Lambda$CDM universe with the same cosmological parameters as in section~\ref{sec:Cl}. We assume that the bias of the bright and faint galaxies evolves as~\cite{biasevol}
\bea
b_{\rm B}(z)=1+(b_{\rm B}^i-1)\frac{D_1(z_i)}{D_1(z)}\, ,\\
b_{\rm F}(z)=1+(b_{\rm F}^i-1)\frac{D_1(z_i)}{D_1(z)}\, ,
\eea
where $b_{\rm B}^i$ and $b_{\rm F}^i$ are the initial values of the bias, for example chosen at redshift $z_i=3$. We choose $b_{\rm B}^i$ and $b_{\rm F}^i$ such that, at $z=0.5$, $b_{\rm B}=2$ and $b_{\rm F}=1.5$.  We neglect magnification bias: $s_{\rm B}=s_{\rm F}=0$. Note that since our calculation is based on linear perturbation theory, the results can not be trusted for $d < 10$\,Mpc.

We see that the amplitude of the monopole and of the quadrupole is significantly larger than the amplitude of the hexadecapole and of the dipole. Measuring the dipole is therefore more challenging. The suppression of the dipole with respect to the monopole or the quadrupole can easily be understood by looking at the form of the functions $\mu_\ell$ and $\nu_\ell$ in equations (\ref{mul}) and (\ref{nul}). The functions $\nu_\ell$  are suppressed by one power of $\HH/k$ with respect to the functions $\mu_\ell$. Hence at subhorizon scales, where $\HH\ll k$, the dipole is smaller than the monopole and the quadrupole: $\xi_1 \sim \HH/k \cdot  \xi_0$.  

However, the amplitude of the dipole may still be large enough to be detected in current galaxy surveys. It is roughly of the same order of magnitude as the hexadecapole, that has already been measured in BOSS (albeit with a low signal-to-noise). Work is in progress to measure the dipole in BOSS. In the future, surveys like Euclid will observe a much larger volume of the sky, and should allow a precise measurement of the dipole. 

The enormous advantage of the multipole expansion over the angular power spectrum expansion is that by fitting for a dipole in the data, one immediately gets rid of the monopole and of the quadrupole contributions. Hence even if these contributions are much larger than the relativistic contributions, they do not spoil the measurement of the dipole. 

Another advantage of the correlation function is that it is quite insensitive to the choice of binning. In figure~\ref{fig:multipole}, no binning has been used. The correlation function has been plotted at fixed separation $d$ and fixed redshift $z$. In practice two types of binning are introduced: first a binning over the position of the bright galaxy, i.e. over $z$, and second a binning over the separation $d$. The effect of the binning in $z$ is very easy to evaluate. It simply introduces an average of the prefactors in equations~(\ref{xistcos}) and~(\ref{xirelcos}). Since all the terms in the prefactors are slowly evolving functions of $z$ (or equivalently of $r$), this average is simply equivalent to evaluating the prefactors at the mean redshift of the bin. The second binning in the separation $d$ is potential more important, since it changes the shape of the functions $\mu_\ell(d)$ defined in equation~(\ref{mul}) and $\nu_\ell(d)$ defined in equation~(\ref{nul}). However, one can show that this change only affects scales smaller than the size of the binning. A binning of a few Mpc has therefore no impact on $d>10$\,Mpc.

\subsubsection{Contaminations to the relativistic dipole}

Equations (\ref{xistcos}) and (\ref{xirelcos}) are valid in the distant-observer approximation. Let us now investigate what happens if we relax this approximation and we calculate the next order in $d/r$. At next order, we expect from equations~(\ref{rp}) to~(\ref{sin}) that the relativistic terms will contribute to the monopole, the quadrupole and the hexadecapole. The amplitude of these contributions will be roughly proportional to $\xi_1\cdot d/r \sim \xi_0\cdot \HH/k \cdot d/r$. They are therefore doubly suppressed (since $\HH\ll k$ and $d\ll r$) with respect to the standard monopole $\xi_0$ and are consequently negligible at all but the largest scales. This is consistent with the conclusion of Yoo and Seljak~\cite{seljak} who show that the relativistic effects have a negligible effect on the monopole, the quadrupole and the hexadecapole.

In addition, at next order we also expect the standard density and redshift distortions to contribute to the dipole and to the octupole. The amplitude of these contributions will be roughly proportional to $\xi_0 \cdot d/r$. Here we see that the suppression $d/r\ll1$ is counterbalanced by the enhancement $k/\HH\gg 1$ in the functions $\mu_\ell$, see~(\ref{mul}). Hence we expect the large-scale corrections from the density and redshift distortions to generate a dipole which is roughly of the same order as the relativistic dipole. It is therefore important to study carefully the form of this term.

Expanding equations (\ref{rp}) to (\ref{sin}) up to next order in $d/r$ and using a Taylor expansion to evaluate the functions $D_1(r'), f(r')$ and $b(r')$ around $r$, we find that the next-order term from equation (\ref{xistalpha}) reads
\bea
\label{stdip}
\fl \xi^{\rm rsd}_{\rm NO}(r, d, \beta)=\frac{2AD_1^2}{9\pi^2\Omega_m^2}\Bigg\{
\Bigg[-(b_{\rm B}-b_{\rm F})\frac{2f}{5}\cdot\mu_2(d)
+(b_{\rm B}-b_{\rm F})\frac{rf'}{6}\Big(\mu_0(d)-\frac{4}{5}\mu_2(d)\Big) \\
-\frac{rf}{6}\big(b'_{\rm B}-b'_{\rm F}\big)\Big(\mu_0(d)-\frac{4}{5}\mu_2(d)\Big)
+\frac{r}{2}\big(b_{\rm B} b'_{\rm F}-b'_{\rm B} b_{\rm F}\big)\cdot\mu_0(d)\Bigg]\cdot\frac{d}{r}\cdot P_1(\cos\beta)\nonumber\\
+\Bigg[(b_{\rm B}-b_{\rm F})\frac{2f}{5} -(b_{\rm B}-b_{\rm F})\frac{rf'}{5}
+\big(b'_{\rm B}-b'_{\rm F}\big)\frac{rf}{5}\Bigg]\mu_2(d)\cdot\frac{d}{r}\cdot P_3(\cos\beta)\Bigg\}\, .\nonumber
\eea

Until now we have neglected the lensing contribution~(\ref{xilens}). We expect however that it will also contribute to the dipole. The bright galaxy is indeed lensed by the faint galaxy only when the bright is behind the faint. And inversely, the faint galaxy is lensed by the bright galaxy only when the faint is behind the bright. Due to the bias difference between the bright and the faint population, this generates an asymmetry in the correlation function. The anti-symmetric lensing contribution can be easily calculated using Limber approximation, and keeping only the lowest order terms in $d/r$. We find
\bea
\xi^{\rm lens}_{\rm anti-symm}=
\frac{A}{6\pi\Omega_m}\frac{D_1^2(r)}{a(r)}\Big[b_{\rm B}(5s_{\rm F}-2)-b_{\rm F}(5s_{\rm B}-2)\Big]\HH_0 \,d\,\cos(\beta)\nonumber\\
\cdot \int_0^\infty \frac{dk_\perp}{k_\perp}\left(\frac{k_\perp}{\HH_0}\right)^3
T^2(k_\perp)J_0\big(k_\perp d\sqrt{1-\cos^2(\beta)}\big)\, .
\eea

\begin{figure}[!t]
\centerline{\epsfig{figure=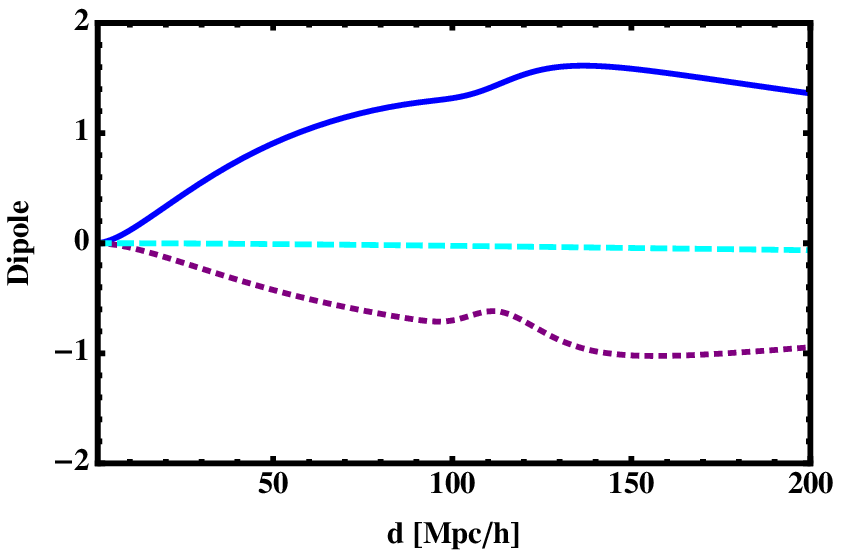,height=5.1cm}\hspace{0.1cm}\epsfig{figure=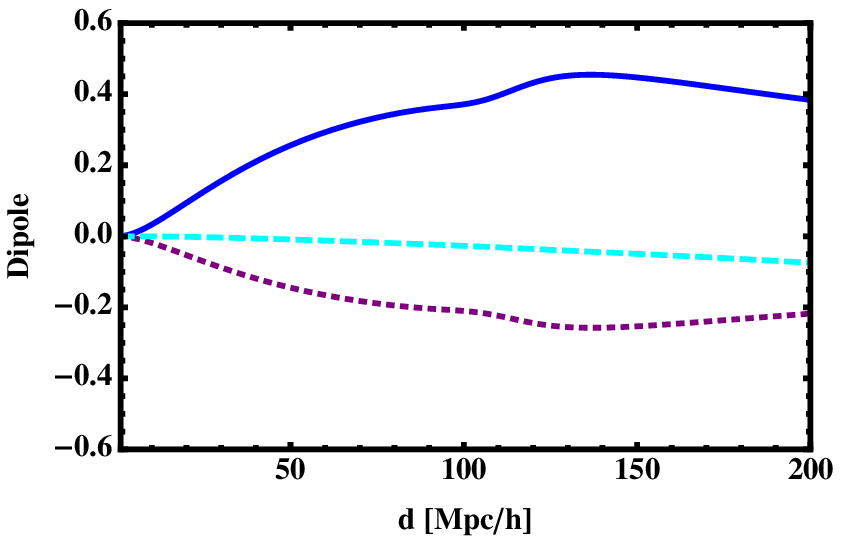,height=5.1cm}}
\caption{ \label{fig:dipole} Amplitude of the dipole (multiplied by $d^2$) at $z=0.25$ (left) and $z=1$ (right) as a function of the separation $d$. In each plot, we show the relativistic contribution (blue solid line) , the lensing contribution (cyan dashed line) and standard redshift-distortion contribution (purple dotted line)}
\end{figure}

In figure~\ref{fig:dipole} we compare the relativistic dipole (\ref{xirelcos}) with the contamination from the standard redshift distortions at next order (\ref{stdip}) and with the gravitational lensing dipole obtained by numerically integrating
\be
\xi^{\rm lens}_1=\frac{3}{2}\int_{-1}^1\!d\mu\, \xi^{\rm lens}_{\rm anti-symm}(r, d, \mu) P_1(\mu)\, .
\ee
We see that the lensing dipole is negligible with respect to the relativistic dipole, except at high redshift and large scales, where its contribution reaches 20\,$\%$. The contamination from the standard redshift distortions at next order on the other hand cannot be neglected, since it is almost of the same amplitude as the relativistic dipole. Hence even though the multipole expansion provides a way of targeting the relativistic effects, it does not allow us to completely get rid of the standard density and redshift  distortions. 

Physically the contributions from the standard terms to the dipole have two origins. First, the evolution with redshift of the density (and more particularly of the bias) and of the velocity of the bright and faint galaxies implies that the number of faint galaxies in front of a bright galaxy is systematically different from the number of faint galaxies behind a bright galaxy.
The second effect is a bit more subtle: it comes from the fact that at large separation, when one exchanges the relative position of the bright and faint galaxy, the angle $\beta$ does not transform as $\beta+\pi$ but as $\beta+\pi-\theta=\alpha+\pi$ (see figure~\ref{fig:cone}). As a consequence, the various terms in equation~(\ref{xistalpha}) are not completely symmetric when the separation $\theta$ is large.

\begin{figure}[!t]
\centerline{\epsfig{figure=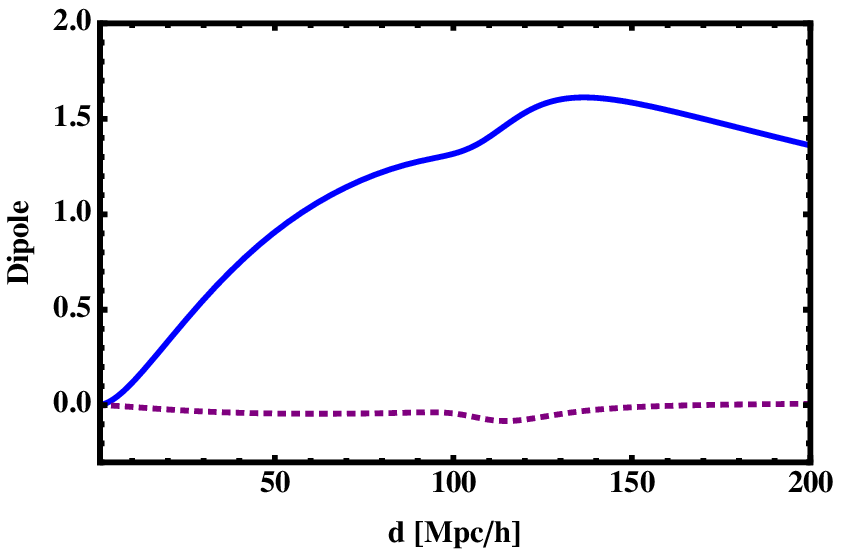,height=5.1cm}\hspace{0.1cm}\epsfig{figure=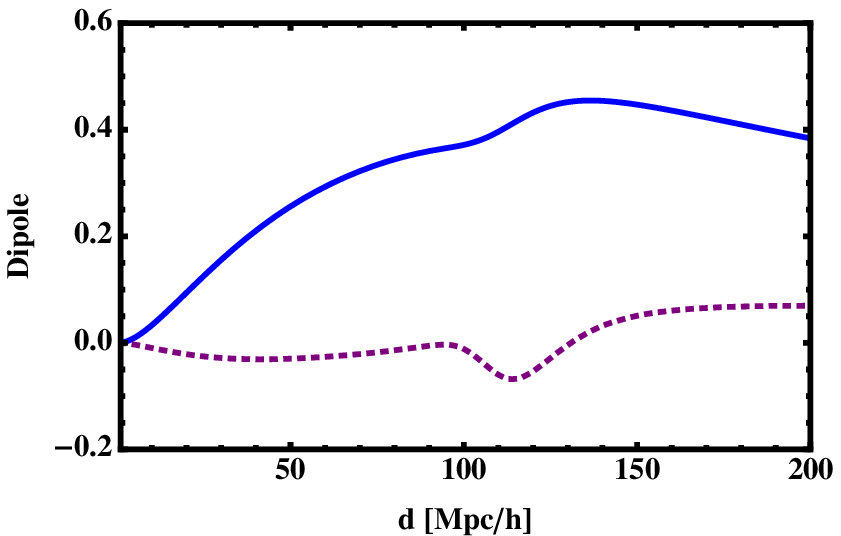,height=5.1cm}}
\caption{ \label{fig:dipolecorr} Amplitude of the dipole (multiplied by $d^2$) at $z=0.25$ (left) and $z=1$ (right) as a function of the separation $d$. The blue solid line is the relativistic contribution and the purple dotted line is the standard redshift-distortion contribution after having removed the wide-angle effect.}
\end{figure}

This {\it wide-angle} effect generates the first term in the first line of equation (\ref{stdip}) and the first term in the last line. The {\it evolution} effect generates all the other terms. Comparing the amplitude of these two types of contributions, we find that the wide-angle effect dominates significantly over the evolution effect~\cite{pap_asymm}. Interestingly, this contribution can be measured and removed from the dipole by combining measurements of the quadrupole of the bright population and of the quadrupole of the faint population. Indeed, we have that
\be
\label{diffquad}
\frac{3d}{10 r}\Big[\xi^{\rm rsd}_{\rm B,\, 2}(r, d, \beta)-\xi^{\rm rsd}_{\rm F,\, 2}(r, d, \beta)\Big]=
-\frac{2A}{9\pi^2\Omega_m^2}\frac{2}{5}(b_{\rm B}-b_{\rm F})D_1^2f\,\mu_2(d)\, ,
\ee
which is exactly equal to the first term in~(\ref{stdip}). In figure~\ref{fig:dipolecorr}, we compare the relativistic dipole with the remaining  standard redshift-space contribution after having removed the wide-angle effect using equation~(\ref{diffquad}). We see that in this case, the contamination from the standard term is negligible at low redshift, and it reaches 10\,$\%$ at redshift 1. 

\section{Conclusion}

In the past few years, it has been shown that relativistic effects distort the observed distribution of galaxies. In this paper, we have reviewed the main steps leading to a fully relativistic expression for the galaxy number counts $\Delta$, valid at linear order. We have shown that besides the standard redshift-space distortions, two types of effects contribute to $\Delta$: Doppler and gravitational redshift effects, that are suppressed by $\HH/k$ with respect to the density and redshift distortions; and gravitational potential effects, that are suppressed by $(\HH/k)^2$. These novel effects are potentially very powerful to test the consistency of general relativity and to constrain modified theories of gravity~\cite{lombriser}, since they depend on the metric potentials $\Phi$ and $\Psi$ and on the peculiar velocity of galaxies $V$, in a different way than the standard Newtonian terms. 

However, measuring these effects in galaxy surveys is difficult, since they are mainly important at large scales, where the number of pairs of galaxies is limited and cosmic variance becomes important. In this paper, we have reviewed two statistical measures that can be used to detect the impact of the relativistic effects: the angular power spectrum, $C_\ell(z,z')$, and the two-point correlation function in configuration space, $\xi(z, d, \beta)$. Even though the angular power spectrum is more closely related to the way observations are performed, the two-point correlation function seems more appropriate to isolate the relativistic effects, since it naturally makes use of the symmetries of the different contributions. Following a novel method presented in~\cite{pap_asymm}, we have shown that the Doppler effect and the gravitational redshift distortions
\be
\Delta^{\rm rel\, V}=\frac{1}{\HH}\dd_r\Psi + \frac{1}{\HH}\dot{\bV}\cdot\bn
-\left[\frac{\dot{\HH}}{\HH^2}+5s-1+\frac{2-5s}{r\HH}\right]\bV\cdot\bn\, , \label{Deltarel}
\ee
generate a dipole modulation in the correlation function. To measure this dipole requires however one refinement: one needs to split the population of galaxies into two populations (for example a bright and a faint population) and to measure the cross-correlation between these populations. 

The huge advantage of this method is that it allows to target the relativistic terms without having to go to very large scales. Fitting for a dipole in the cross-correlation function allows indeed to get rid of most of the standard redshift-space distortions contributions, which strongly dominate at small scales. 

By looking at equation~(\ref{Deltarel}), we see that the Doppler effect and the gravitational redshift distortions allow an immediate model-independent test of gravity. In theories of gravitation that obey Euler equation, the combination in~(\ref{Deltarel}) reduces indeed to
 \be
\Delta^{\rm rel\, V}=-\left[\frac{\dot{\HH}}{\HH^2}+5s+\frac{2-5s}{r\HH}\right]\bV\cdot\bn\, . \label{DeltaEuler}
\ee
By comparing the dipole with the monopole and the quadrupole contributions in~(\ref{xistcos}), one can therefore test the validity of equation~(\ref{DeltaEuler}) and consequently probe the relationship between the velocity and the gravitational potential $\Psi$.  
Relativistic effects in galaxy surveys offer consequently exciting prospects to test the fundamental nature of gravity at cosmological scales. A careful forecast of the signal-to-noise of these effects in future surveys, like Euclid or WFIRST, and of the sensitivity of these terms to various modified gravity models remains to be made.   

Let us mention finally that the relativistic effects contribute not only to the galaxy number counts and to the convergence as discussed here, but also to other observables, like the luminosity distance of supernovae~\cite{pap_dL}, the cosmic shear~\cite{pap_shear} and the 21\,cm intensity mapping~\cite{pap_21cm}. Combining these different observables in the relativistic regime has the potential to place new constraints on cosmological models. Furthermore, the work presented here is valid in the linear regime. In the non-linear regime, a large number of new relativistic terms contribute to large-scale structure observations. Careful derivations of the second-order contributions to the shear~\cite{pap_shear}, the convergence~\cite{conv_2nd} and the galaxy number counts~\cite{delta_2nd} have been performed recently, opening the possibility of interesting new tests of gravity. 

\ack

I thank Ruth Durrer, Enrique Gaztanaga and Lam Hui, who co-authored the two main papers~\cite{pap_gal} and \cite{pap_asymm} on which this review is based and Marc-Olivier Bettler for his help with the figures. I am supported by King's College Cambridge.

\appendix

\section{Coefficients of the standard redshift-space correlation function, $S_i$}
\label{app:coeffst}

The standard coefficients of equation~(\ref{xistalpha}) are
\bea
\fl S_1(r,r',d)=D_1(r)D_1(r')\Bigg[\left(b_{\rm B}(r)b_{\rm F}(r')+\frac{b_{\rm B}(r)}{3}f(r')+\frac{b_{\rm F}(r')}{3}f(r)+\frac{2}{15}f(r)f(r')\right)\mu_0(d)\nonumber\\
-\frac{1}{3}\left(\frac{b_{\rm B}(r)}{2}f(r')+\frac{b_{\rm F}(r')}{2}f(r)+\frac{2}{7}f(r)f(r') \right)\mu_2(d) +\frac{3}{140}f(r)f(r')\mu_4(d)\Bigg]\, ,\nonumber\\
\fl S_2(r,r',d)=-D_1(r)D_1(r')\Bigg[\left(\frac{b_{\rm F}(r')}{2}f(r)+\frac{3}{14}f(r)f(r') \right)\mu_2(d) +\frac{1}{28}f(r)f(r')\mu_4(d)\Bigg]\, ,\nonumber\\
\fl S_3(r,r',d)=-D_1(r)D_1(r')\Bigg[\left(\frac{b_{\rm B}(r)}{2}f(r')+\frac{3}{14}f(r)f(r') \right)\mu_2(d) +\frac{1}{28}f(r)f(r')\mu_4(d)\Bigg]\, ,\nonumber\\
\fl S_4(r,r',d)=D_1(r)D_1(r')f(r)f(r')\left[\frac{1}{15}\mu_0(d)-\frac{1}{21}\mu_2(d)+\frac{19}{140}\mu_4(d) \right]\, ,\nonumber\\
\fl S_5(r,r',d)=D_1(r)D_1(r')f(r)f(r')\left[\frac{1}{15}\mu_0(d)-\frac{1}{21}\mu_2(d)-\frac{4}{35}\mu_4(d) \right]\, .\nonumber
\eea

\section{Coefficients of the relativistic correlation function, $R_i$}
\label{app:coeffrel}

The standard coefficients of equation~(\ref{xirelalpha}) are

\bea
\fl R_1(r,r',d)=G_{\rm F}(r')D_1(r)D_1(r')f(r')
\left[\left(b_{\rm B}(r)+\frac{2}{5}f(r) \right)\nu_1(d)-\frac{1}{10}f(r)\nu_3(d) \right]\, ,\nonumber\\
\fl R_2(r,r',d)=-G_{\rm B}(r)D_1(r)D_1(r')f(r)
\left[\left(b_{\rm F}(r')+\frac{2}{5}f(r') \right)\nu_1(d)-\frac{1}{10}f(r')\nu_3(d) \right]\, ,\nonumber\\
\fl R_3(r,r',d)=G_{\rm F}(r')D_1(r)D_1(r')f(r)f(r')\frac{1}{5}\left[\nu_1(d)-\frac{3}{2}\nu_3(d) \right]\, ,\nonumber\\
\fl R_4(r,r',d)=-G_{\rm B}(r)D_1(r)D_1(r')f(r)f(r')\frac{1}{5}\left[\nu_1(d)-\frac{3}{2}\nu_3(d)\right]\, ,\nonumber\\
\fl R_5(r,r',d)=G_{\rm F}(r')D_1(r)D_1(r')f(r)f(r')\frac{1}{5}\Big[\nu_1(d)+\nu_3(d) \Big]\, ,\nonumber\\
\fl R_6(r,r',d)=-G_{\rm B}(r)D_1(r)D_1(r')f(r)f(r')\frac{1}{5}\Big[\nu_1(d)+\nu_3(d)\Big]\, ,\nonumber
\eea
where the functions $G_{\rm F}$ and $G_{\rm B}$ are defined as
\bea
G_{\rm F}(r') = \left[\frac{\dot{\HH}(r')}{\HH^2(r')}+\frac{2}{r'\HH(r')}+5s_{\rm F}(r')\left(1-\frac{1}{r'\HH(r')}\right)\right]\frac{\HH(r')}{\HH_0}\, ,\nonumber\\
G_{\rm B}(r) = \left[\frac{\dot{\HH}(r)}{\HH^2(r)}+\frac{2}{r\HH(r)}+5s_{\rm B}(r)\left(1-\frac{1}{r\HH(r)}\right)\right]\frac{\HH(r)}{\HH_0}\, .\nonumber
\eea

\end{document}